\documentclass{article}
\usepackage{spconf,amsmath,graphicx,bm,standalone,tikz,hyperref,pgfplots}
\usepackage[caption=false,font=footnotesize,labelfont=sf,textfont=sf]{subfig}
\usetikzlibrary{calc, trees, positioning, arrows, shapes, fit, backgrounds, patterns}
\pgfplotsset{compat=newest}

\graphicspath{{./imgs/}}

\def\L{{\cal L}}

\title{On Annotation-free Optimization of Video Coding for Machines}
\name{Marc Windsheimer, Fabian Brand, Andr\'e Kaup\thanks{The authors gratefully acknowledge that this work has been funded by the Deutsche Forschungsgemeinschaft (DFG, German Research Foundation) under project number 426084215.}}
\address{Multimedia Communications and Signal Processing\\Friedrich-Alexander-Universit\"at Erlangen-N\"urnberg\\Cauerstr. 7, 91058 Erlangen, Germany}
\newcommand\copyrighttext{%
	\footnotesize \textcopyright 2024 IEEE. Personal use of this material is permitted. Permission from IEEE must be obtained for all other uses, in any current or future media, including reprinting/republishing this material for advertising or promotional purposes, creating new collective works, for resale or redistribution to servers or lists, or reuse of any copyrighted component of this work in other works.
	
}
\newcommand\copyrightnoticeOwn{%
	\begin{tikzpicture}[remember picture,overlay]
		\node[anchor=north,yshift=-10pt] at (current page.north) {\fbox{\parbox{\dimexpr\textwidth-\fboxsep-\fboxrule\relax}{\copyrighttext}}};
	\end{tikzpicture}%
	\vspace{-8mm}
}

\begin{document}
	%\ninept
	%
	\maketitle
	\copyrightnoticeOwn
	\begin{abstract}
	Today, image and video data is not only viewed by humans, but also automatically analyzed by computer vision algorithms. 
	However, current coding standards are optimized for human perception. 
	Emerging from this, research on video coding for machines tries to develop coding methods designed for machines as information sink. 
	Since many of these algorithms are based on neural networks, most proposals for video coding for machines build upon neural compression. 
	So far, optimizing the compression by applying the task loss of the analysis network, for which ground truth data is needed, is achieving the best coding performance. 
	But ground truth data is difficult to obtain and thus an optimization without ground truth is preferred. 
	In this paper, we present an annotation-free optimization strategy for video coding for machines.
	We measure the distortion by calculating the task loss of the analysis network.
	Therefore, the predictions on the compressed image are compared with the predictions on the original image, instead of the ground truth data.
	Our results show that this strategy can even outperform training with ground truth data with rate savings of up to 7.5 \%.
	By using the non-annotated training data, the rate gains can be further increased up to 8.2 \%.
	\end{abstract}
	\begin{keywords}
		Video Coding for Machines, Neural Network Compression, Computer Vision, Learned Image Coding, Machine-to-Machine Communication
	\end{keywords}
	\section{Introduction}
	\label{sec:intro}
	Machine-to-machine communication is one of the main causes for the growing amount of data transmitted over the internet.
	By far the most significant contributions to global data traffic are related to image and video data \cite{sandvine2023}.
	The automated analysis of transmitted image and video data is one application of M2M communication.
	However, image codecs, like JPEG or BPG, and video codecs, like \mbox{HEVC \cite{sullivan2012}} or VVC \cite{bross2021}, are designed for the perceptual characteristics of the human visual system.
	These codecs compress the visual data in a lossy procedure.
	Thus, not only the redundancy is decreased, but also deviations between the coded and the original data are introduced.
	For traditional image and video coding, irrelevancy reduction removes parts of the original data, for which the human visual system is less sensitive.
	A typical example for this are high-frequency components.
	Since the computer vision algorithms have different characteristics compared to the human visual system, the relevancy of certain image details varies as well.
	As machine-to-machine communication gains importance, novel coding schemes specifically optimized for machines and algorithms as information sink are required.
	As a result, an ad-hoc group on \mbox{\textit{Video Coding for Machines (VCM)} \cite{zhang2019}} has been introduced by MPEG in 2019.
	Their goal is to standardize a bit stream format optimized for machine-to-machine scenarios.
	
	Adapting neural compression for VCM scenarios usually requires annotated and ideally pristine, i.e. uncoded, image and video data.
	However, availability of this type of data is limited and annotating images or videos is time-consuming and associated with high costs.
	Therefore, there is a demand of annotation-free optimization methods for VCM scenarios.
	
	In this paper, we focus on improving the training framework for VCM-optimized neural image compression following the compress-then-analyze paradigm \cite{redondi2016}.
	Our approach does not require annotated data and achieves even better coding results when compared with VCM optimization with available ground truth data.

	\section{Neural Image Compression for Machines}
	\label{sec:related}
	\subsection{Neural Image Compression}
	\begin{figure*}
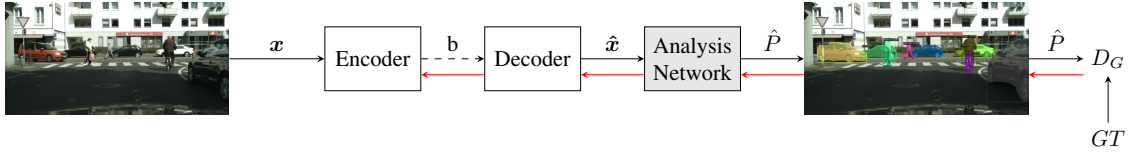

		\centering
		\includestandalone[scale=.85]{img/real_strategy}
		\caption{Traditional training strategy with ground truth data. Red arrows denote gradient flow during backpropagation. Weights in shaded blocks are not updated during optimization.}
		\label{fig:real_strategy}
	\end{figure*}
	\begin{figure*}
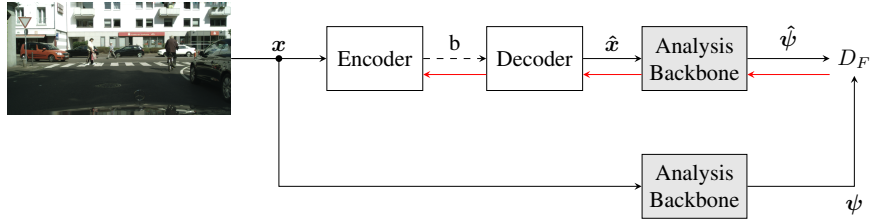

		\centering
		\includestandalone[scale=.85]{img/feat_strategy}
		\caption{Training strategy with feature loss. Red arrows denote gradient flow during backpropagation. Weights in shaded blocks are not updated during optimization.}
		\label{fig:feat_strategy}
	\end{figure*}
	The research field of image compression is currently dominated by compressive autoencoders \cite{balle2017}.
	They form an end-to-end trainable neural network, which transforms the original image $\bm{x}$ into a latent space representation $\bm{y}$.
	The latent space is quantized to $\bm{\hat{y}}$, entropy coded into the \mbox{bit stream $b$} and transmitted to the decoding device.
	The decoding transform then reconstructs the image $\bm{\hat{x}}$ from the quantized latent space. 
	
	During the training process, the quantization is replaced by an addition with uniform noise.
	The rate-distortion loss $\L_{RD}$ for the optimization of a compressive autoencoder consists of an estimate of the bit rate $R$ and a distortion measure $D(\bm{\hat{x}}, \bm{x})$.
	When the compression network is optimized for the human visual system, common distortion metrics are MSE or MS-SSIM \cite{wang2003}.
	To weight the bit rate and the distortion metric, a trade-off parameter $\lambda$ can be defined, resulting in the following loss function:
	\begin{equation}
		\L_{RD} = R + \lambda D(\bm{\hat{x}}, \bm{x})\text{.}
		\label{eq:cae_loss}
	\end{equation}
	Multiple models at different rate points can be trained by varying the parameter $\lambda$.
	
	\subsection{VCM Optimization with Ground Truth Data}
	In order to optimize neural compression for VCM scenarios, other distortion metrics have to be used.
	A common approach is to borrow the loss of the analysis task \cite{choi2022}.
	Here, the reconstructed image $\bm{\hat{x}}$ is fed into the task network $T$ and the resulting predictions $\hat{P}=T(\bm{\hat{x}})$ are compared with the ground truth data $GT$.
	The distortion metric $D_G$ with ground truth data is then calculated as:
	\begin{equation}
		D_G(\bm{\hat{x}}, GT) = \L_{TASK}(T(\bm{\hat{x}}), GT)\text{,}
		\label{eq:task_loss}
	\end{equation}
	where $\L_{TASK}$ represents the loss used for the optimization of the analysis network.
	The resulting gradients flow through the frozen analysis network into the compression network.
	This type of training strategy is depicted in Fig. \ref{fig:real_strategy}.
	
	However, as annotated data is required, the amount of training data is typically limited.
	Moreover, since erroneous predictions of the task network may even occur on uncompressed data, a potentially significant portion of the resulting task loss is not related to the compression performance but to the general shortcomings of the task network itself.

	\subsection{VCM Optimization with Feature-based Losses}
	One solution for the VCM scenario is to train the compression network to generate images which result in similar predictions on the compressed data as they would be on the original pristine data. 
	Such an approach is the use of feature-based metrics \cite{fischer2020,fischer2022b} for optimization.
	As depicted in Fig. \ref{fig:feat_strategy}, the initial layers of the backbone $f_B$ of the task network are used to calculate intermediary features $\bm{\psi}=f_B(\bm{x})$ and $\bm{\hat{\psi}}=f_B(\bm{\hat{x}})$ of the original and the reconstructed image, respectively.
	The feature-based distortion $D_F$ is then derived by calculating the difference of both features, e.g. via the SSE:
	\begin{equation}
		D_F(\bm{\hat{\psi}}, \bm{\psi}) = \sum (\bm{\hat{\psi}} - \bm{\psi})^2\text{.}
		\label{eq:feature_loss}
	\end{equation}
	A major advantage of this method is that it does not require annotated training data.
	Therefore, this loss can be used to train on non-annotated data, which is easier to obtain, or to adapt the compression network for a different domain.
	However, the coding results in \cite{fischer2022b} show that models trained with a feature-based loss can not reach the performance of models trained with task loss using ground truth data.
	Thus, there is still a need for a VCM loss, which works on non-annotated data and matches or surpasses the performance of the task loss with ground truth.

	\section{Annotation-free Optimization of Image Coding for Machines}
	\label{sec:main}
	\begin{figure*}
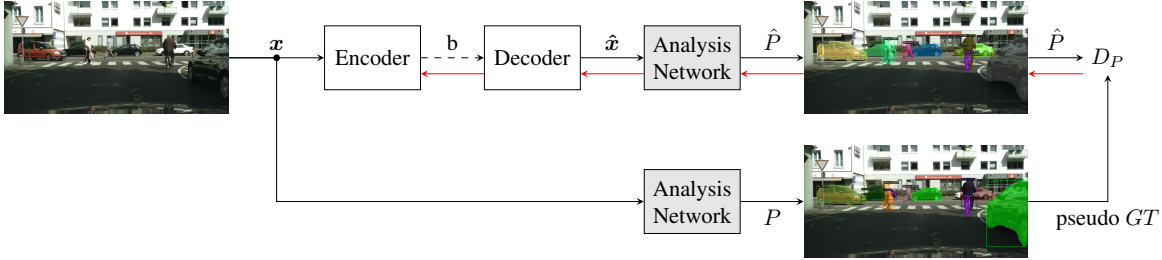

		\centering
		\includestandalone[scale=.85]{img/strategy}
		\caption{Proposed training strategy with pseudo ground truth. Red arrows denote gradient flow during backpropagation. Weights in shaded blocks are not updated during optimization.}
		\label{fig:strategy}
	\end{figure*}
	In this paper, we propose to use the predictions of the task model on the uncompressed image as pseudo ground truth for the task loss during VCM optimization.
	So far, pseudo ground truth has only been used for the evaluation of VCM codecs on unlabeled data \cite{fischer2022a}.
	Harell et al. \cite{harell2022} showed that performing feature matching in deeper layers of the analysis network leads to a better compression performance in the VCM scenario.
	Our approach can be seen as a continuation of this insight.
	Instead of evaluating intermediate features, we compare the final predictions.
	An overview of our training strategy is shown in Fig. \ref{fig:strategy}.
	The image $\bm{x}$ is compressed in a lossy way by the image compression network, resulting in the reconstruction $\bm{\hat{x}}$.
	The reconstructed image is analyzed by the task network to obtain the predictions $\hat{P}=T(\bm{\hat{x}})$.
	Furthermore, the predictions $P=T(\bm{x})$ on the original image are derived.
	The distortion $D_P$ is calculated by comparing the \mbox{predictions $\hat{P}$} on the reconstructed frame $\bm{\hat{x}}$ with the pseudo ground truth.
	The distortion measure (\ref{eq:task_loss}) is then modified as follows:
	\begin{equation}
		D_P(\bm{\hat{x}}, \bm{x}) = \L_{TASK}(T(\bm{\hat{x}}), T(\bm{x}))\text{.}
		\label{eq:pseudo_task_loss}
	\end{equation}
	Identical to the training with ground truth data, the obtained gradients are backpropagated through the analysis network into the compression network.
	
	Using pseudo ground truth has two major advantages.
	First, it does not require any annotations.
	Thus, the strategy can be used to optimize compression networks with a larger amount of training data, which could otherwise not be used due to a lack of annotations.
	Second, it focuses the loss on the direct influence of compression artifacts on the task performance.
	Since the analysis networks are not free of errors, even on uncompressed data, parts of the loss using ground truth data is not related to the performance of the compression network but due to the limitations of the analysis network instead.
	Therefore, these errors can hardly be corrected by the compression network.
	In contrast, when the task loss is derived with respect to the pseudo ground truth, the whole loss is related to the deviation between the reconstructed and the original image.
	Hence, it is a more reliable measure to calculate the coding distortion for the VCM scenario.
	
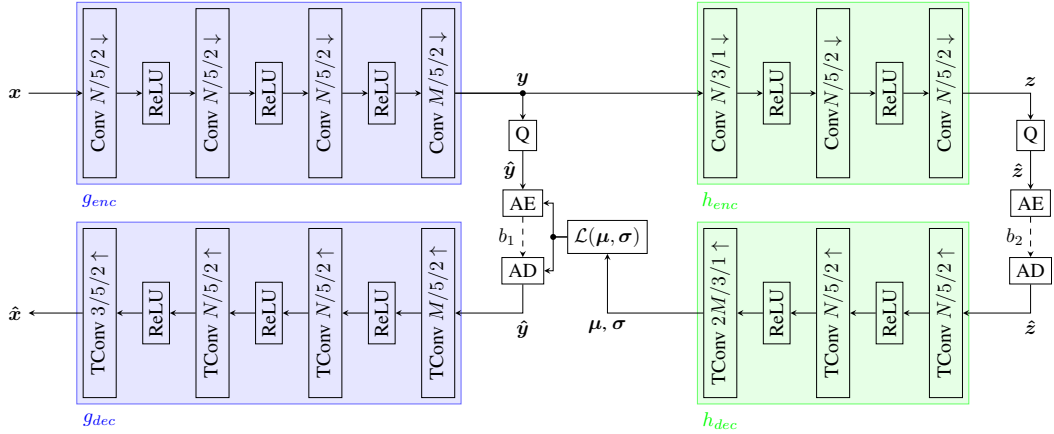
\begin{figure*}
		\centering
			\begin{tikzpicture}[
		dataBlob/.style={rectangle, anchor = center},
		Layer/.style={rectangle, draw, anchor=center, minimum width=3cm, minimum height=0.5cm},
		Misc/.style={rectangle, draw, anchor = center},
		Code/.style={, draw, anchor = center, fill=black!20},
		bg/.style={rectangle, anchor = center, fill=\colorBG, anchor = north west,},
		myArrow/.style={-stealth},
		label/.style={rectangle, anchor = center},
		operator/.style={circle ,draw, anchor = center, minimum size = 0.2cm, inner sep = 0pt},
		dot/.style={circle, fill, minimum size=3pt, inner sep=0pt}
		]
		
		\node (xt) [dataBlob, align=center] at (0, 0) {$\bm{x}$};
		\node (gec1) [Layer, right of=xt, node distance=1.5cm, rotate=90] {Conv $N/5/2\downarrow$};
		\node (ger1) [Misc, right of=gec1, rotate=90] {ReLU};
		\node (gec2) [Layer, right of=ger1, rotate=90] {Conv $N/5/2\downarrow$};
		\node (ger2) [Misc, right of=gec2, rotate=90] {ReLU};
		\node (gec3) [Layer, right of=ger2, rotate=90] {Conv $N/5/2\downarrow$};
		\node (ger3) [Misc, right of=gec3, rotate=90] {ReLU};
		\node (gec4) [Layer, right of=ger3, rotate=90] {Conv $M/5/2\downarrow$};
		\node (lge) [label, below of=gec1, node distance=1.9cm, text=blue] {$g_{enc}$};
		
		\begin{pgfonlayer}{background}
			\node (shade) [draw=blue!60, fill=blue!10, fit=(gec1) (gec4), inner sep=3pt] {};
		\end{pgfonlayer}
		
		\node (hec1) [Layer, right of=gec4, node distance=5cm, rotate=90] {Conv $N/3/1\downarrow$};
		\node (her1) [Misc, right of=hec1, rotate=90] {ReLU};
		\node (hec2) [Layer, right of=her1, rotate=90] {Conv$N/5/2\downarrow$};
		\node (her2) [Misc, right of=hec2, rotate=90] {ReLU};
		\node (hec3) [Layer, right of=her2, rotate=90] {Conv $N/5/2\downarrow$};
		\node (lhe) [label, below of=hec1, node distance=1.9cm, text=green] {$h_{enc}$};
		
		\begin{pgfonlayer}{background}
			\node (shade) [draw=green!60, fill=green!10, fit=(hec1) (hec3), inner sep=3pt] {};
		\end{pgfonlayer}
		
		\node (hq) [Misc, right of=hec3, node distance=1.5cm, yshift=-0.75cm] {Q};
		\node (hae) [Misc, below of=hq, node distance=1.2cm] {AE};
		\node (had) [Misc, below of=hae, node distance=1.2cm] {AD};
		
		\node (gq) [Misc, right of=gec4, node distance=1.5cm, yshift=-0.75cm] {Q};
		\node (gae) [Misc, below of=gq, node distance=1.2cm] {AE};
		\node (gad) [Misc, below of=gae, node distance=1.2cm] {AD};
		
		\node (laplace) [Misc, right of=gae, node distance=1.5cm, yshift=-0.6cm] {$\mathcal{L}(\bm{\mu}, \bm{\sigma})$};
		
		\node (gdc4) [Layer, left of=gad, node distance=1.5cm, yshift=-0.75cm, rotate=90] {TConv $M/5/2\uparrow$};
		\node (gdr3) [Misc, left of=gdc4, rotate=90] {ReLU};
		\node (gdc3) [Layer, left of=gdr3, rotate=90] {TConv $N/5/2\uparrow$};
		\node (gdr2) [Misc, left of=gdc3, rotate=90] {ReLU};
		\node (gdc2) [Layer, left of=gdr2, rotate=90] {TConv $N/5/2\uparrow$};
		\node (gdr1) [Misc, left of=gdc2, rotate=90] {ReLU};
		\node (gdc1) [Layer, left of=gdr1, rotate=90] {TConv $3/5/2\uparrow$};
		\node (xthat) [dataBlob, left of=gdc1, node distance=1.5cm] {$\bm{\hat{x}}$};
		\node (lgd) [label, below of=gdc1, node distance=1.9cm, text=blue] {$g_{dec}$};
		
		\begin{pgfonlayer}{background}
			\node (shade) [draw=blue!60, fill=blue!10, fit=(gdc1) (gdc4), inner sep=3pt] {};
		\end{pgfonlayer}
		
		\node (hdc1) [Layer, right of=gdc4, node distance=5cm, rotate=90] {TConv $2M/3/1\uparrow$};
		\node (hdr1) [Misc, right of=hdc1, rotate=90] {ReLU};
		\node (hdc2) [Layer, right of=hdr1, rotate=90] {TConv $N/5/2\uparrow$};
		\node (hdr2) [Misc, right of=hdc2, rotate=90] {ReLU};
		\node (hdc3) [Layer, right of=hdr2, rotate=90] {TConv $N/5/2\uparrow$};
		\node (lhd) [label, below of=hdc1, node distance=1.9cm, text=green] {$h_{dec}$};
		
		\begin{pgfonlayer}{background}
			\node (shade) [draw=green!60, fill=green!10, fit=(hdc1) (hdc3), inner sep=3pt] {};
		\end{pgfonlayer}
	
		\draw[myArrow] (xt) -- (gec1);
		\draw[myArrow] (gec1) -- (ger1);
		\draw[myArrow] (ger1) -- (gec2);
		\draw[myArrow] (gec2) -- (ger2);
		\draw[myArrow] (ger2) -- (gec3);
		\draw[myArrow] (gec3) -- (ger3);
		\draw[myArrow] (ger3) -- (gec4);
		\draw[myArrow] (gec4) -- (hec1);
		\draw[myArrow] (hec1) -- (her1);
		\draw[myArrow] (her1) -- (hec2);
		\draw[myArrow] (hec2) -- (her2);
		\draw[myArrow] (her2) -- (hec3);
		
		\draw[myArrow] (hec3) -| (hq) node[pos=0.5, above] {$\bm{z}$};
		\draw[myArrow] (hq) -- (hae) node[pos=0.5, left] {$\bm{\hat{z}}$};
		\draw[myArrow, dashed] (hae) -- (had) node[pos=0.5, left] {$b_2$};
		\draw[myArrow] (had) |- (hdc3) node[pos=0.5, below] {$\bm{\hat{z}}$};
		
		\draw[myArrow] (gec4) -| (gq) node[pos=0.5, dot] {} node[pos=0.5, above] {$\bm{y}$};
		\draw[myArrow] (gq) -- (gae) node[pos=0.5, left] {$\bm{\hat{y}}$};
		\draw[myArrow, dashed] (gae) -- (gad) node[pos=0.5, left] {$b_1$};
		\draw[myArrow] (gad) |- (gdc4) node[pos=0.5, below] {$\bm{\hat{y}}$};
		
		\draw[myArrow] (hdc3) -- (hdr2);
		\draw[myArrow] (hdr2) -- (hdc2);
		\draw[myArrow] (hdc2) -- (hdr1);
		\draw[myArrow] (hdr1) -- (hdc1);
		\draw[myArrow] (hdc1) -| (laplace) node[pos=0.5, below] {$\bm{\mu}$, $\bm{\sigma}$};
		\draw[myArrow] (gdc4) -- (gdr3);
		\draw[myArrow] (gdr3) -- (gdc3);
		\draw[myArrow] (gdc3) -- (gdr2);
		\draw[myArrow] (gdr2) -- (gdc2);
		\draw[myArrow] (gdc2) -- (gdr1);
		\draw[myArrow] (gdr1) -- (gdc1);
		\draw[myArrow] (gdc1) -- (xthat);
		
		\draw[myArrow] (laplace.west) -- ++(-0.25,0) |- (gae) node[pos=0, dot] {};
		\draw[myArrow] (laplace.west) -- ++(-0.25,0) |- (gad);
		
	\end{tikzpicture}
		\caption{Detailed overview of the autoencoder used for performing the experiments. It consists of the core autoencoder (blue) and the hyperprior coder (green). Conv $c/k/s\downarrow$ and TConv $c/k/s\uparrow$ denote a convolutional and transposed convolutional layer, respectively, with the number of output channels $c$, the kernel size $k$ and stride $s$. $\L$ denotes a Laplacian distribution with mean $\bm{\mu}$ and scale $\bm{\sigma}$. $b_1$ and $b_2$ represent the bit streams required for transmitting the latent representation $\bm{\hat{y}}$ and the hyperprior $\bm{\hat{z}}$, respectively.}
		\label{fig:coder}
	\end{figure*}

	We validate our annotation-free training strategy with an image compression autoencoder similar to \cite{balle2018}.
	However, we replace the zero-mean Gaussian model of the latent space with a Laplacian distribution with mean and scale parameter, as this has shown better performance in \cite{zhou2018}.
	According to \cite{fischer2021}, we replace the activations with rectified linear units \mbox{(ReLUs)}.
	The architecture of the coder is depicted in \mbox{Fig. \ref{fig:coder}}.
	Since the applied compression network, based on an autoencoder with hyperprior, shares a similar structure with most neural image compression networks, it can be assumed that the results of the different training strategies for VCM can also be transferred to other image compression approaches.
	
	\section{Experiments}
	\label{sec:experiments}
	We investigate the tasks of instance segmentation and semantic segmentation on the Cityscapes \cite{cordts2016} dataset.
	The dataset consists of 2975 RGB images in the training set and additional 500 images in the validation set.
	For each image, pixel-accurate masks are available.
	Each labeled image is extracted from a 30 frame long video sequence.
	For the first task, we apply a Mask R-CNN \cite{he2017} network with feature-pyramid \mbox{structure \cite{lin2017}} from the model zoo of the \mbox{Detectron2}\footnote{\url{https://github.com/facebookresearch/detectron2}} library.
	For semantic segmentation we use a pretrained \mbox{DeepLabV3+ \cite{chen2018}} model with MobileNetv2 \cite{sandler2018} backbone\footnote{\url{https://github.com/VainF/DeepLabV3Plus-Pytorch}}.
	We train and compare separate image compression networks for both tasks using ground truth data and our proposed pseudo ground truth.
	As additional reference, we also obtained models trained with feature-based losses.
	For Mask R-CNN, the feature loss is calculated in the 'p2' feature space in the feature pyramid, whereas for DeepLabV3+, we compare the high-level features generated from the MobileNetv2 backbone.
	Furthermore, VVC intra coding from the reference software VTM\footnote{\url{https://vcgit.hhi.fraunhofer.de/jvet/VVCSoftware_VTM}} in version 20.2 is applied. 
	
	\subsection{Training}
	Before the models are finetuned on the VCM task, we first pretrain the network for 125.000 iterations using MSE loss as distortion metric.
	We then obtain multiple task-optimized models at different bit rates by switching the distortion metric to the task loss using either ground truth data or pseudo ground truth.
	For the Mask R-CNN models, we set the rate points to $\lambda=[16, 8, 4, 2]$ and for DeepLabV3+ to $\lambda=[64, 32, 16, 8]$.
	We finetune the networks for the VCM task for ten epochs each.
	Additionally, we trained models using the feature-based loss as proposed in \cite{fischer2022b}.
	Here, we set all corresponding values for the trade-off parameter $\lambda$ to obtain similar rate points.
	
	\subsection{Evaluation}
	To evaluate the different training losses, we compress the 500 pristine validation images of the Cityscapes dataset with all models.
	We calculate the bit rate in bits per pixel (bpp) and the traditional image metrics PSNR and MS-SSIM.
	Additionally, we obtain the weighted average precision (wAP) \cite{fischer2020b} for the models optimized for Mask R-CNN and the mean intersection over union (mIOU) for the DeepLabV3+ models with respect to the ground truth annotations.

	\section{Results}
	\label{sec:results}
	\subsection{Main experiment}
	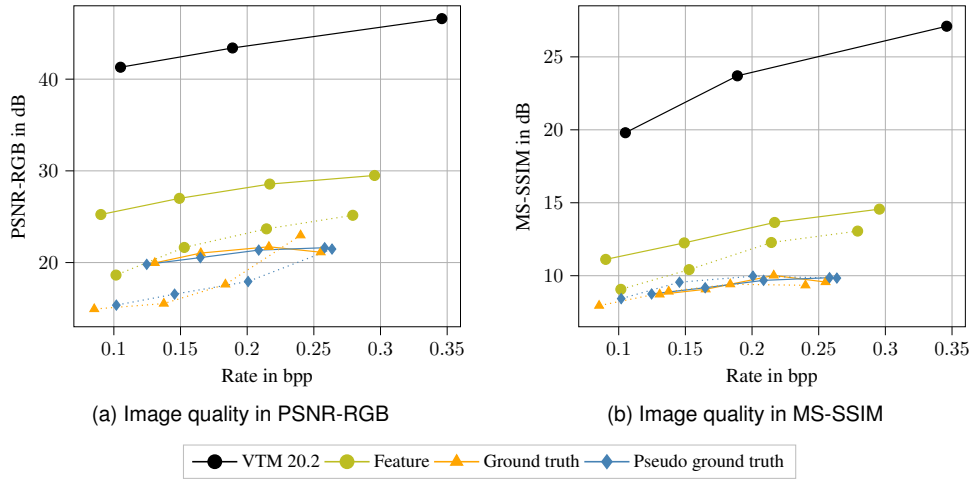
\begin{figure*}
		\centering
		\subfloat[Image quality in PSNR-RGB]{% This file was created with tikzplotlib v0.10.1.
\begin{tikzpicture}

\definecolor{darkgray176}{RGB}{176,176,176}
\definecolor{goldenrod18818934}{RGB}{188,189,34}
\definecolor{orange}{RGB}{255,165,0}
\definecolor{steelblue}{RGB}{70,130,180}

\begin{axis}[
tick align=outside,
tick pos=left,
x grid style={darkgray176},
xlabel={Rate in bpp},
xmajorgrids,
xmin=0.07, xmax=0.36,
xtick style={color=black},
y grid style={darkgray176},
ylabel={PSNR-RGB in dB},
ymajorgrids,
ymin=13, ymax=48,
ytick style={color=black}
]
\addplot [semithick, black, mark=*, mark size=2.5, mark options={solid}]
table {%
0.105 41.3
0.189 43.4
0.346 46.6
};
\addplot [semithick, goldenrod18818934, mark=*, mark size=2.5, mark options={solid}]
table {%
0.29554411315918 29.497809337616
0.216928298950195 28.5582644615173
0.149164245605469 27.0017588500977
0.0902503814697266 25.2404040374756
};
\addplot [semithick, orange, mark=triangle*, mark size=2.5, mark options={solid}]
table {%
0.255147705078125 21.1294864730835
0.216281585693359 21.7207952728272
0.165211120605469 21.0355840568542
0.130711181640625 19.9625126609802
};
\addplot [semithick, steelblue, mark=diamond*, mark size=2.5, mark options={solid}]
table {%
0.258119125366211 21.6187989311218
0.208655624389648 21.3605461273193
0.164861053466797 20.5416428146362
0.124658874511719 19.8051410160065
};
\addplot [semithick, goldenrod18818934, dotted, mark=*, mark size=2.5, mark options={solid}]
table {%
0.279201568603516 25.1552538223267
0.214473220825195 23.6763839874268
0.152815979003906 21.6483900146484
0.101594055175781 18.6367265892029
};
\addplot [semithick, orange, dotted, mark=triangle*, mark size=2.5, mark options={solid}]
table {%
0.240060791015625 22.9752876815796
0.18367204284668 17.6163202514648
0.13748974609375 15.5268929214478
0.0852911987304688 14.9320703449249
};
\addplot [semithick, steelblue, dotted, mark=diamond*, mark size=2.5, mark options={solid}]
table {%
0.263635940551758 21.4724003677368
0.200700927734375 17.9361029262543
0.145553848266602 16.5570216197968
0.101866882324219 15.3582579021454
};
\end{axis}

\end{tikzpicture}}
		\hspace{10pt}
		\subfloat[Image quality in MS-SSIM]{% This file was created with tikzplotlib v0.10.1.
\begin{tikzpicture}

\definecolor{darkgray176}{RGB}{176,176,176}
\definecolor{goldenrod18818934}{RGB}{188,189,34}
\definecolor{orange}{RGB}{255,165,0}
\definecolor{steelblue}{RGB}{70,130,180}

\begin{axis}[
tick align=outside,
tick pos=left,
x grid style={darkgray176},
xlabel={Rate in bpp},
xmajorgrids,
xmin=0.07, xmax=0.36,
xtick style={color=black},
y grid style={darkgray176},
ylabel={MS-SSIM in dB},
ymajorgrids,
ymin=6.5, ymax=28.5,
ytick style={color=black}
]
\addplot [semithick, black, mark=*, mark size=2.5, mark options={solid}]
table {%
0.105 19.8
0.189 23.7
0.346 27.1
};
\addplot [semithick, goldenrod18818934, mark=*, mark size=2.5, mark options={solid}]
table {%
0.29554411315918 14.5619048333168
0.216928298950195 13.6500051426888
0.149164245605469 12.2501202178001
0.0902503814697266 11.1180707550049
};
\addplot [semithick, orange, mark=triangle*, mark size=2.5, mark options={solid}]
table {%
0.255147705078125 9.56589811444283
0.216281585693359 10.0183910477161
0.165211120605469 9.0690191578865
0.130711181640625 8.71819867610931
};
\addplot [semithick, steelblue, mark=diamond*, mark size=2.5, mark options={solid}]
table {%
0.258119125366211 9.86467022895813
0.208655624389648 9.68123845219612
0.164861053466797 9.17619961023331
0.124658874511719 8.7485286962986
};
\addplot [semithick, goldenrod18818934, dotted, mark=*, mark size=2.5, mark options={solid}]
table {%
0.279201568603516 13.0578587889671
0.214473220825195 12.2763098335266
0.152815979003906 10.4170010316372
0.101594055175781 9.06503167271614
};
\addplot [semithick, orange, dotted, mark=triangle*, mark size=2.5, mark options={solid}]
table {%
0.240060791015625 9.34216318249702
0.18367204284668 9.41638069391251
0.13748974609375 8.91636037111282
0.0852911987304688 7.94710215330124
};
\addplot [semithick, steelblue, dotted, mark=diamond*, mark size=2.5, mark options={solid}]
table {%
0.263635940551758 9.8371545124054
0.200700927734375 9.9676973760128
0.145553848266602 9.55425802826881
0.101866882324219 8.42752505540848
};
\end{axis}

\end{tikzpicture}}\\
		\vspace{-5pt}
		\subfloat{\begin{tikzpicture}

\definecolor{darkgray176}{RGB}{176,176,176}
\definecolor{goldenrod18818934}{RGB}{188,189,34}
\definecolor{orange}{RGB}{255,165,0}
\definecolor{steelblue}{RGB}{70,130,180}

\begin{axis}[%
	hide axis,
	xmin=10,
	xmax=50,
	ymin=0,
	ymax=0.4,
	legend columns=4,
	legend style={nodes={scale=0.7, transform shape}, draw=lightgray,legend cell align=left}
	]
	\addlegendimage{semithick, black, mark=*, mark size=2.5, mark options={solid}}
	\addlegendentry{VTM 20.2};
	
	\addlegendimage{semithick, goldenrod18818934, mark=*, mark size=2.5, mark options={solid}}
	\addlegendentry{Feature};
	
	\addlegendimage{semithick, orange, mark=triangle*, mark size=2.5, mark options={solid}}
	\addlegendentry{Ground truth};
	
	\addlegendimage{semithick, steelblue, mark=diamond*, mark size=2.5, mark options={solid}}
	\addlegendentry{Pseudo ground truth};	
\end{axis}

\end{tikzpicture}}
		\caption{Coding results for image metrics PSNR and MS-SSIM averaged over the 500 Cityscapes validation images. Solid lines represent models trained for Mask R-CNN, whereas dotted lines symbolize the DeepLabV3+ models.}
		\label{fig:trad}
	\end{figure*}
	\begin{figure*}
		\centering
		\subfloat[Task performance of models optimized for instance segmentation with Mask R-CNN]{\begin{tikzpicture}

\definecolor{darkgray}{RGB}{169,169,169}
\definecolor{darkgray176}{RGB}{176,176,176}
\definecolor{goldenrod18818934}{RGB}{188,189,34}
\definecolor{orange}{RGB}{255,165,0}
\definecolor{steelblue}{RGB}{70,130,180}

\begin{axis}[
tick align=outside,
tick pos=left,
x grid style={darkgray176},
xlabel={Rate in bpp},
xmajorgrids,
xmin=0.07, xmax=0.36,
xtick style={color=black},
y grid style={darkgray176},
ylabel={Weighted AP in \%},
ymajorgrids,
ymin=30, ymax=40,
ytick style={color=black}
]
\addplot [semithick, darkgray, dashed]
table {%
0 39.6071211422057
1 39.6071211422057
};
\addplot [semithick, black, mark=*, mark size=2.5, mark options={solid}]
table {%
0.105 30.5
0.189 35.2
0.346 36.1
};
\addplot [semithick, goldenrod18818934, mark=*, mark size=2.5, mark options={solid}]
table {%
0.29554411315918 34.8654541860235
0.216928298950195 34.6602154788969
0.149164245605469 34.4142235840664
0.0902503814697266 32.7367895621232
};
\addplot [semithick, orange, mark=triangle*, mark size=2.5, mark options={solid}]
table {%
0.255147705078125 36.2839478106158
0.216281585693359 36.2293071068499
0.165211120605469 35.1416526638331
0.130711181640625 34.5070277750321
};
\addplot [semithick, steelblue, mark=diamond*, mark size=2.5, mark options={solid}]
table {%
0.258119125366211 36.4895917762182
0.208655624389648 36.2760600968667
0.164861053466797 35.3658792132055
0.124658874511719 34.5670752199268
};
\end{axis}

\end{tikzpicture}}
		\hspace{10pt}
		\subfloat[Task performance of models optimized for semantic segmentation with DeepLabV3+]{\begin{tikzpicture}

\definecolor{darkgray}{RGB}{169,169,169}
\definecolor{darkgray176}{RGB}{176,176,176}
\definecolor{goldenrod18818934}{RGB}{188,189,34}
\definecolor{orange}{RGB}{255,165,0}
\definecolor{steelblue}{RGB}{70,130,180}

\begin{axis}[
tick align=outside,
tick pos=left,
x grid style={darkgray176},
xlabel={Rate in bpp},
xmajorgrids,
xmin=0.07, xmax=0.36,
xtick style={color=black},
y grid style={darkgray176},
ylabel={Mean IOU in \%},
ymajorgrids,
ymin=45, ymax=75,
ytick style={color=black}
]
\addplot [semithick, darkgray, dashed]
table {%
0 72.0676
1 72.0676
};
\addplot [semithick, black, mark=*, mark size=2.5, mark options={solid}]
table {%
0.105 49.1
0.189 59.7
0.346 66.2
};
\addplot [semithick, goldenrod18818934, dotted, mark=*, mark size=2.5, mark options={solid}]
table {%
0.279201568603516 64.2589
0.214473220825195 62.959
0.152815979003906 61.123
0.101594055175781 56.9102
};
\addplot [semithick, orange, dotted, mark=triangle*, mark size=2.5, mark options={solid}]
table {%
0.240060791015625 66.5837
0.18367204284668 66.1331
0.13748974609375 64.8885
0.0852911987304688 62.2715
};
\addplot [semithick, steelblue, dotted, mark=diamond*, mark size=2.5, mark options={solid}]
table {%
0.263635940551758 66.9316
0.200700927734375 66.2362
0.145553848266602 65.5623
0.101866882324219 63.7841
};
\end{axis}

\end{tikzpicture}}\\
		\vspace{-5pt}
		\subfloat{\begin{tikzpicture}

\definecolor{darkgray176}{RGB}{176,176,176}
\definecolor{goldenrod18818934}{RGB}{188,189,34}
\definecolor{orange}{RGB}{255,165,0}
\definecolor{steelblue}{RGB}{70,130,180}

\begin{axis}[%
	hide axis,
	xmin=10,
	xmax=50,
	ymin=0,
	ymax=0.4,
	legend columns=4,
	legend style={nodes={scale=0.7, transform shape}, draw=lightgray,legend cell align=left}
	]
	\addlegendimage{semithick, black, mark=*, mark size=2.5, mark options={solid}}
	\addlegendentry{VTM 20.2};
	
	\addlegendimage{semithick, goldenrod18818934, mark=*, mark size=2.5, mark options={solid}}
	\addlegendentry{Feature};
	
	\addlegendimage{semithick, orange, mark=triangle*, mark size=2.5, mark options={solid}}
	\addlegendentry{Ground truth};
	
	\addlegendimage{semithick, steelblue, mark=diamond*, mark size=2.5, mark options={solid}}
	\addlegendentry{Pseudo ground truth};	
\end{axis}

\end{tikzpicture}}
		\caption{Coding results for task metrics weighted average precision and mean intersection over union averaged over the 500 Cityscapes validation images. Solid lines represent models trained for Mask R-CNN, whereas dotted lines symbolize the DeepLabV3+ models. Dashed grey line corresponds to task performance on uncompressed data.}
		\label{fig:task}
	\end{figure*}
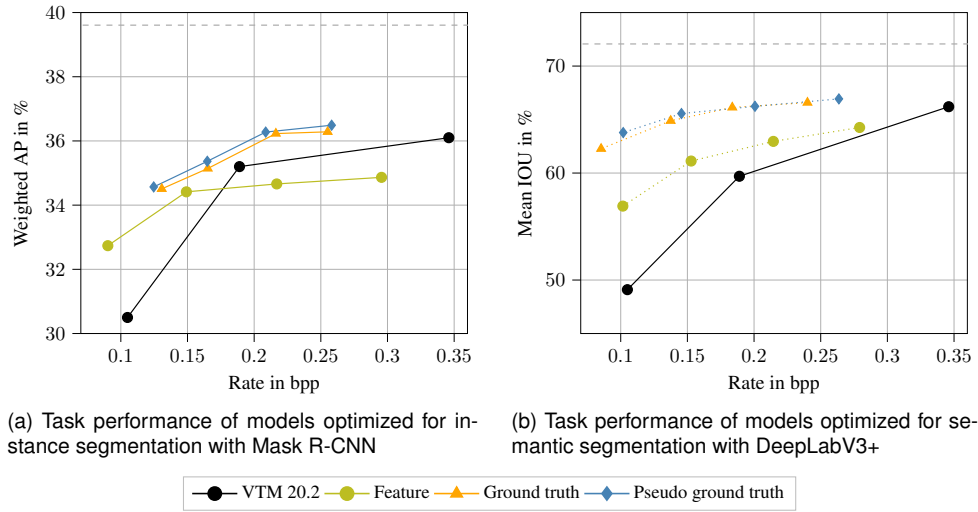
	
	The coding results for the different training strategies are shown in Figs. \ref{fig:trad} and \ref{fig:task}.
	In terms of the traditional image quality metrics PSNR and MS-SSIM, see Fig. \ref{fig:trad}, the feature-based loss clearly produces reconstructions closer to the uncompressed images as compared to both the task loss with ground truth and pseudo ground truth.
	The reason for this is that the feature-based loss is closer to the pixel space and therefore leads to more similar results in terms of traditional pixel-oriented metrics like MSE or MS-SSIM.
	However, for the task-related metrics, as can be seen in Fig. \ref{fig:task} the feature loss models can not reach the performance of the task loss models.
	The Bj{\o}ntegaard delta (BD) values in Table \ref{tab:bdvalues} show that both VVC and the feature loss models can not achieve the same compression performance of the task loss with ground truth data.
	The task loss with pseudo ground truth, however, can accomplish even better coding performances compared to the task loss with ground truth data.
	Calculating the Bj{\o}ntegaard delta rates (BDR) results in a noteworthy rate reduction of \mbox{5.1 \%} for wAP and \mbox{7.5 \%} for mIOU.
	\begin{table}
		\caption{Bj{\o}ntegaard delta quality in percentage points (pp) for the VCM metrics using the model trained with ground truth data as anchor. BD + metric denotes the average difference in percentage points (pp) in the corresponding metric for the overlapping bit rate range. Higher is better. Best values are set in bold.\label{tab:bdvalues}}
		\centering
		\vspace{10pt}
		\begin{tabular}{l|ccc}
			\hline
					& VTM 20.2	 & Feature	& Pseudo GT\\
			\hline
			BD wAP	&	-0.73 pp &	-1.00 pp	&	\textbf{0.18 pp}\\
			BD mIOU	& 	-8.73 pp &	-4.44 pp	&	\textbf{0.23 pp}\\
			\hline
		\end{tabular}
	\end{table}

	\subsection{Benefit of Optimization with Pseudo Ground Truth}
	\begin{figure*}
		\centering
		\subfloat[Task performance as wAP of task loss models optimized for instance segmentation with Mask R-CNN]{\begin{tikzpicture}

\definecolor{darkgray}{RGB}{169,169,169}
\definecolor{darkgray176}{RGB}{176,176,176}
\definecolor{firebrick}{RGB}{178,34,34}
\definecolor{orange}{RGB}{255,165,0}
\definecolor{steelblue}{RGB}{70,130,180}

\begin{axis}[
tick align=outside,
tick pos=left,
x grid style={darkgray176},
xlabel={Rate in bpp},
xmajorgrids,
xmin=0.07, xmax=0.36,
xtick style={color=black},
y grid style={darkgray176},
ylabel={Weighted AP in \%},
ymajorgrids,
ymin=34, ymax=40,
ytick style={color=black}
]
\addplot [semithick, darkgray, dashed]
table {%
0 39.6071211422057
1 39.6071211422057
};
\addplot [semithick, orange, mark=triangle*, mark size=2.5, mark options={solid}]
table {%
0.255147705078125 36.2839478106158
0.216281585693359 36.2293071068499
0.165211120605469 35.1416526638331
0.130711181640625 34.5070277750321
};
\addplot [semithick, steelblue, mark=diamond*, mark size=2.5, mark options={solid}]
table {%
0.258119125366211 36.4895917762182
0.208655624389648 36.2760600968667
0.164861053466797 35.3658792132055
0.124658874511719 34.5670752199268
};
\addplot [semithick, firebrick, mark=pentagon*, mark size=2.5, mark options={solid}]
table {%
0.259462417845937 36.8344667391519
0.204353671258025 36.3184343184739
0.160861053466797 35.3681130769991
0.129658874511719 34.6982998912721
};
\end{axis}

\end{tikzpicture}}
		\hspace{10pt}
		\subfloat[Task performance as mIOU of task loss models optimized for semantic segmentation with DeepLabV3+]{\begin{tikzpicture}

\definecolor{darkgray}{RGB}{169,169,169}
\definecolor{darkgray176}{RGB}{176,176,176}
\definecolor{firebrick}{RGB}{178,34,34}
\definecolor{orange}{RGB}{255,165,0}
\definecolor{steelblue}{RGB}{70,130,180}

\begin{axis}[
tick align=outside,
tick pos=left,
x grid style={darkgray176},
xlabel={Rate in bpp},
xmajorgrids,
xmin=0.07, xmax=0.36,
xtick style={color=black},
y grid style={darkgray176},
ylabel={Mean IOU in \%},
ymajorgrids,
ymin=61.5, ymax=72.5,
ytick style={color=black}
]
\addplot [semithick, darkgray, dashed]
table {%
0 72.0676
1 72.0676
};
\addplot [semithick, orange, dotted, mark=triangle*, mark size=2.5, mark options={solid}]
table {%
0.240060791015625 66.5837
0.18367204284668 66.1331
0.13748974609375 64.8885
0.0852911987304688 62.2715
};
\addplot [semithick, steelblue, dotted, mark=diamond*, mark size=2.5, mark options={solid}]
table {%
0.263635940551758 66.9316
0.200700927734375 66.2362
0.145553848266602 65.5623
0.101866882324219 63.7841
};
\addplot [semithick, firebrick, dotted, mark=pentagon*, mark size=2.5, mark options={solid}]
table {%
0.253548110961914 66.9445
0.206939071655273 66.6568
0.144073150634766 65.3853
0.104985641479492 64.371
};
\end{axis}

\end{tikzpicture}}\\
		\vspace{-5pt}
		\subfloat{\begin{tikzpicture}
	
	\definecolor{darkgray176}{RGB}{176,176,176}
	\definecolor{goldenrod18818934}{RGB}{188,189,34}
	\definecolor{orange}{RGB}{255,165,0}
	\definecolor{steelblue}{RGB}{70,130,180}
	\definecolor{firebrick}{RGB}{178,34,34}
	
	\begin{axis}[%
		hide axis,
		xmin=10,
		xmax=50,
		ymin=0,
		ymax=0.4,
		legend columns=3,
		legend style={nodes={scale=0.7, transform shape}, draw=lightgray,legend cell align=left}
		]	
		\addlegendimage{semithick, orange, mark=triangle*, mark size=2.5, mark options={solid}}
		\addlegendentry{Ground truth};
		
		\addlegendimage{semithick, steelblue, mark=diamond*, mark size=2.5, mark options={solid}}
		\addlegendentry{Pseudo ground truth};
		
		\addlegendimage{semithick, firebrick, mark=pentagon*, mark size=2.5, mark options={solid}}
		\addlegendentry{Pseudo ground truth + random frame};	
	\end{axis}
	
\end{tikzpicture}}
		\caption{Coding results for task metrics for task loss models. Solid lines represent models trained for Mask R-CNN, whereas dotted lines symbolize the DeepLabV3+ models. Dashed grey line corresponds to task performance on uncompressed data.}
		\label{fig:benefit}
	\end{figure*}
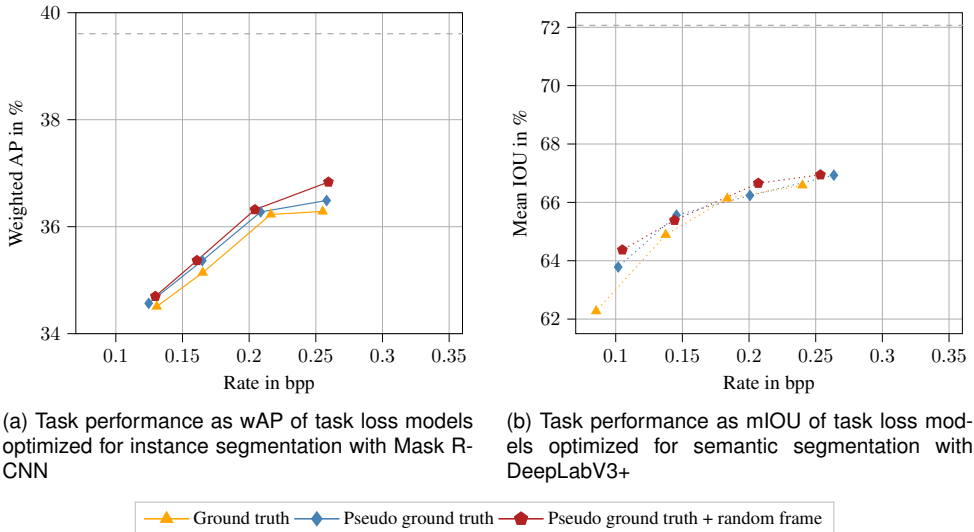
	In the previous section, we compared the different training strategies using only the labeled frame within each 30 frame long sequence.
	However, since our annotation-free optimization strategy using pseudo ground truth does not rely on labels, we can use the whole sequence data.
	Therefore, we retrained the compression model using pseudo ground truth.
	At each step, one frame is randomly selected from the corresponding video sequence, which is essentially a temporal augmentation of the previously used training data.
	By taking a random frame from each sequence instead of directly adding all available frames to the training data, we  train for the same amount of steps and thus, ensure a fair comparison.
	In \mbox{Fig. \ref{fig:benefit}}, we depict the coding results for the task metrics of this approach in comparison with the ground truth data and pseudo ground truth calculated on the labeled validation frames.
	The temporal augmentation can further improve the coding performance of the model with pseudo ground truth.
	As shown in Table \ref{tab:bdvalues2}, we obtain BD gains of up to 0.34 pp over training with ground truth data and \mbox{0.12 pp} with respect to pseudo ground truth without temporal augmentation.
	This highlights one of the key benefits of pseudo ground truth over ground truth data.
	Since annotations are no longer required, a larger dataset can be used to optimize the compression networks for the VCM scenario.
	\begin{table}
		\caption{Bj{\o}ntegaard delta quality in percentage points (pp) and Bj{\o}ntegaard delta rate in \% for the VCM metrics with the task loss with and without temporal augmentation using the model trained with ground truth data as anchor. BD + metric denotes the average difference in percentage points (pp) in the corresponding metric for the overlapping bit rate range. Higher is better. BDR + metric denotes the average rate difference in \% for equivalent task performance. Lower is better. Best values are set in bold.\label{tab:bdvalues2}}
		\centering
		\vspace{10pt}
		\begin{tabular}{l|cc}
			\hline
			& Pseudo GT		& Pseudo GT\\
			& labeled frame	& random frame\\
			\hline
			BD wAP	&	0.18 pp	&	\textbf{0.30 pp}\\
			BD mIOU	&	0.23 pp	&	\textbf{0.34 pp}\\
			\hline
			BDR wAP	&	-5.1 \% &	\textbf{-6.8 \%}\\
			BDR mIOU&	-7.5 \% &	\textbf{-8.2 \%}\\
			\hline
		\end{tabular}
	\end{table}

	\section{Conclusion}
	\label{sec:conclusion}
	In this paper, we proposed the use of the prediction results of computer vision algorithms on uncompressed images as pseudo ground truth for the optimization of neural image compression for machines.
	This method allows to train neural compression for VCM scenarios on non-annotated image data.
	We evaluated the approach by comparing it with the training on ground truth data and a strategy, which measures the similarity in the feature space for estimating the distortion.
	
	The conducted experiments prove that pseudo ground truth can be successfully used for optimizing neural image compression with respect to computer vision tasks.
	Since the task loss with pseudo ground truth only focuses on the differences between the prediction results on the compressed and original frames, the real impact of the compression is evaluated.
	In contrast, the loss using ground truth data additionally contains the present errors of the task model.
	Therefore, only a fraction of the calculated loss is related to the performance of the compression network.
	Our method achieves average rate savings of about 5.1 \% with respect to wAP on Mask R-CNN and 7.5 \% for mIOU on DeepLabV3+.
	Since our method does not require annotated training data, we can further improve the performance by using a larger amount of training data.
	With this enlarged training set, we obtain even higher rate savings of 6.8 \% for wAP and 8.2 \% for mIOU.
	
	As annotated and pristine video data is hardly available, this method is especially suited for the training of neural video compression in VCM scenarios.
	Therefore, future research can investigate whether similar gains can be reached when applying pseudo ground truth for the optimization of neural video compression.

	% Below is an example of how to insert images. Delete the ``\vspace'' line,
	% uncomment the preceding line ``\centerline...'' and replace ``imageX.ps''
	% with a suitable PostScript file name.
	% -------------------------------------------------------------------------

	%\vfill
	%\pagebreak

	%\vfill\pagebreak
	
	\bibliographystyle{IEEEbib}
	\bibliography{references}
	
\end{document}